\documentclass[a4paper]{article}
\usepackage{amsmath}
\usepackage{graphicx, color}
\usepackage{authblk}

\title{Correlated insulating states in slow Dirac fermions on a honeycomb moir{\'e}  superlattice}
\author[1,2,$\dagger$]{Dongyang Yang}
\author[1,2,$\dagger$]{Jing Liang}
\author[1,2,$\dagger$]{Haodong Hu}
\author[1,2]{Nitin Kaushal}
\author[3,4]{Chih-En Hsu}
\author[5]{Kenji Watanabe}
\author[6]{Takashi Taniguchi}
\author[1,2]{Jerry. I Dadap}
\author[4]{Zhenglu Li}
\author[1,2]{Marcel Franz}
\author[1,2,*]{Ziliang Ye}

\affil[1]{Department of Physics and Astronomy, The University of British Columbia, Vancouver, BC V6T 1Z1, Canada}
\affil[2]{Quantum Matter Institute, The University of British Columbia, Vancouver, BC V6T 1Z4, Canada}
\affil[3]{Department of Physics, Tamkang University, New Taipei City 251301, Taiwan}
\affil[4]{Mork Family Department of Chemical Engineering and Materials Science, University of Southern California, Los Angeles, California 90089, USA}
\affil[5]{Research Center for Functional Materials, National Institute for Materials Science, 1-1 Namiki, Tsukuba 305-0044, Japan}
\affil[6]{International Center for Materials Nanoarchitectonics, National Institute for Materials Science,  1-1 Namiki, Tsukuba 305-0044, Japan}
\affil[$\dagger$]{These authors contribute equally to this work}
\affil[*]{Corresponding author: zlye@phas.ubc.ca}
\date{}

\begin{document}
\maketitle
	
\begin{abstract}
\textbf{Strong Coulomb repulsion is predicted to open a many-body charge gap at the Dirac point of graphene, transforming the semimetal into a Mott insulator. However, this correlated insulating phase has remained inaccessible in pristine graphene, where a large Fermi velocity dominates the interaction effects. To overcome this limitation, we realize a honeycomb moir{\'e} superlattice in a twisted MoSe$_2$ homobilayer, where a graphene-like band structure forms with a Fermi velocity reduced by nearly two orders of magnitude. These slow moir{\'e} bands are folded from the valence band maximum at the $\Gamma$ valley of the extended Brillouin zone with negligible spin-orbital coupling, and can therefore simulate massless Dirac fermions in the strongly correlated regime with full SU(2) symmetry. By correlating Rydberg exciton sensing with moir{\'e} trions of different spatial characters, we detect a Mott gap at the Dirac point that persists up to 110 K. We further identify correlated insulating states at $\nu=-1$ with a weak ferromagnetic coupling as well as at several fractional fillings. Our results highlight the potential of studying a wide range of quantum many-body phenomena in twisted two-dimensional materials.}
\end{abstract}

\section*{Main}
\subsection*{Introduction}
The physics of massless Dirac fermions has been extensively studied in monolayer graphene, whose two-dimensional honeycomb lattice hosts low-energy excitations with linear energy dispersion at the corners of the Brillouin zone ($\pm$ K valleys)\cite{novoselov2005two}. Near the Dirac point where the conduction and valence bands meet, the electron's pseudospin (sublattice index) is collinear with its momentum, giving rise to opposite chiralities in the two valleys\cite{katsnelson2006chiral}. Theoretically, if the Coulomb interaction between Dirac electrons is sufficiently large, the material is predicted to transform from the semimetal to a Mott insulator through a Gross-Neveu transition\cite{sorella1992semi,herbut2006interactions,herbut2009theory}. In the Mott insulating phase, an energy gap opens at the Dirac point and the chiral symmetry becomes broken\cite{semenoff2012chiral,triola2015many, gutierrez2016imaging}, giving rise to a range of exotic phenomena, including quantum spin liquids\cite{meng2010quantum}, topological superconductivity\cite{nandkishore2012chiral}, and non-Fermi-liquid behavior\cite{gonzalez1994non}. However, such a correlated insulating state remains unobserved in pristine graphene since the Coulomb interaction between electrons is too weak compared to their hopping energy, or, in other words, the Fermi velocity is too large\cite{kotov2012electron}. Here we utilize the honeycomb moir{\'e} superlattice in a twisted MoSe$_2$ homobilayer to experimentally simulate slow Dirac fermions in the strong correlation regime. 

Transition metal dichalcogenide (TMD) homo- and heterostructures have emerged as a powerful platform for exploring exotic physics with non-trivial topology and strong correlations\cite{tang2020simulation,zeng2023thermodynamic,cai2023signatures,park2023observation,mak2022semiconductor}. Among these, the twisted MoSe$_2$ homobilayer (t-MoSe$_2$) is an ideal system for studying slow Dirac fermion physics for two reasons\cite{angeli2021gamma,zhang2021electronic}. First, the moir{\'e} potential minima arising from the interlayer coupling are distributed symmetrically between the MX and XM sites (MX denotes the stacking where the Mo atom is on top of the Se atom and vice versa), forming the same honeycomb lattice as graphene (Figure 1a \& b). Second, the highest filled moir{\'e} bands are folded from the valence band maximum (VBM) in the $\Gamma$ valley of the monolayer Brillouin zone, where all states are doubly spin-degenerate following the Kramers' theorem (Figure 1c)\cite{fang2015ab}. As a result, a twisted MoSe$_2$ homobilayer shares qualitatively the same single-particle effective Hamiltonian near the Dirac point as a monolayer graphene, except that the Fermi velocity becomes over thirty times slower. Unlike in folded K valley bands, the wavefunction in the moir{\'e} bands folded from the $\Gamma$-valley VBM has substantial interlayer overlaps, leading to a strong layer-hybridized character \cite{liang2022optically,foutty2023tunable,yang2024non}.

Since the hopping energy ($t$) between MX and XM sites increases more rapidly than the on-site Coulomb repulsion ($U_0$) with twist angle, the strong correlation phenomena ($U_0>t$) must be explored in the small-twist-angle limit. Nevertheless, if the twist angle becomes too small, significant atomic relaxation can occur, reducing the moir{\'e} potential and on-site Coulomb energy\cite{sung2020broken,andersen2021excitons}. Here we focus on t-MoSe$_2$ homobilayers with twist angles between 2$^\circ$ and 4$^\circ$, where the moir{\'e} superlattice constant, a$_M$, is comparable to the AB-BA domain wall width, ensuring that the lattice reconstruction is sufficiently weak to preserve the Coulomb repulsion strength. Using the Bistritzer-MacDonald continuum model, we calculate the single-particle band structure of an ideal 3.6$^{\circ}$ t-MoSe$_2$ bilayer (Figure 1d), where the Dirac point appears at the filling of two holes per moire unit cell ($\nu=-2$). The top two valence bands (shown in red) are sufficiently isolated from the other bands so that we can map them onto a single-orbital Hubbard model with a $t$ about 5 meV and $U_0$ of 100-200 meV\cite{angeli2021gamma}. In this large $U_0/t$ limit, a variety of correlated states with novel magnetic orders have been predicted\cite{meng2010quantum,kaushal2022magnetic}. Here we probe these states optically through the 2s exciton in a nearby sensor layer, as well as the moir{\'e} trions in t-MoSe$_2$ with different spatial characters.

\subsection*{Probing insulating states in t-MoSe$_2$ homobilayer}

Building on the optical studies on TMD heterostructures\cite{xu2020correlated}, we employ the Rydberg exciton sensing technique to investigate the correlated states in a t-MoSe$_2$ homobilayer with a targeted twist angle of 3.6$^{\circ}$, fabricated using the dry transfer method. In addition to fabricating the device in a dual gate geometry that can independently control doping and electric field ($E$), we placed a WS$_2$ monolayer 1 nm (the thickness of a trilayer hBN)  from the t-MoSe$_2$ as an exciton sensor (Figure 1e and Extended Figure 1). Due to the type-II band alignment between WS$_2$ and MoSe$_2$ (Figure 1c), holes are primarily doped into the MoSe$_2$ layer, allowing precise tuning of the filling factor ($\nu$) in the moir{\'e} superlattice. If the t-MoSe$_2$ enters an insulating/incompressible state, the screening of the WS$_2$ 2s exciton is reduced. (In contrast to the 1s state, the 2s exciton has a Bohr radius of about 5 nm and is therefore highly sensitive to the dielectric environment\cite{zhang2022correlated}.) By monitoring the peak energy and oscillator strength of the WS$_2$ 2s exciton, we can optically probe the electronic compressibility of the t-MoSe$_2$ homobilayer.

Figure 1f shows the first energy derivative of the reflection contrast spectrum, $d(\Delta R/R)/d\hbar \omega$, of the device under the intrinsic condition. The measurement is performed at 7 K to reduce nonlinear gating effects. The two absorption peaks at approximately 1.610 and 1.812 eV correspond to the A and B exciton resonances of the moir{\'e} MoSe$_2$ bilayer. The strongest feature, centered at 2.072 eV, arises from the 1s exciton in the WS$_2$ sensor layer, while the 2s exciton peak appears at 2.192 eV. The narrow linewidths of the WS$_2$ excitons confirm that the energy and charge transfer between the sensor and sample layers are effectively suppressed by the hBN spacer. 

Since our MoSe$_2$ flake is chemically n-doped, the Fermi level of the system can be tuned through the intrinsic band gap by applying an outward electric field symmetrically between the top and bottom gates. Throughout this tuning, we continuously monitor the 2s exciton in the sensor layer with its doping dependence shown in Figure 2b. At about $V_{tg}=-1$ V, we observe an abrupt 18-meV blueshift of the 2s exciton resonance and a large enhancement in its oscillator strength. (We define the oscillator strength as the peak-to-peak amplitude of the resonance, as shown in Extended Figure 2.) According to the exciton screening picture, this enhancement indicates that the sample layer has reached the charge-neutral condition (with an initial doping level of $\sim$10$^{11}$ cm$^{-2}$). Further increasing the gate voltage moves the Fermi level into the valence band, whose maximum resides in the $\Gamma$ valley as we will show below.

In the low-doping region near $\nu=0$, we observe a peak splitting in the 2s exciton resonance, which we attribute to the formation of interlayer attractive and repulsive Fermi polarons (IAP and IRP) arising from the Coulomb interaction between excitons in the WS$_2$ layer and free carriers in the MoSe$_2$ layer\cite{cui2024interlayer}. As the carrier density increases, the IRP gradually transfers its oscillator strength to the IAP\cite{huang2023quantum}. Interestingly, within the gate voltage range $-4<V_{tg}<-1$ V, we observe a series of blueshifts in the IAP peak, accompanied by an enhancement of its oscillator strength. These behaviors indicate a reduced screening and the emergence of insulating states at these doping levels. 

We attribute the two distinct blueshifts observed at $V_{tg}=-2.36$ V and $V_{tg}=-3.48$ V to the two integer fillings ($\nu=-1$ and $\nu=-2$) for two reasons. First, the magnitudes of these blueshifts are notably larger than those at other fillings. Second, only at these two fillings do the blueshifts persist at elevated temperatures, as we will show in the next section. Since correlated insulating states at integer fillings are expected to exhibit larger charge gaps\cite{xu2020correlated,zhang2022correlated}, we associate them with the fillings of one and two holes per moir{\'e} cell, respectively. From the voltage separation between these two features, we extract the moir{\'e} cell density, $n_M \approx 4.6 \times10^{12}cm^{-2}$, which corresponds to a twist angle of 3.8$^{\circ}$ (or 5.0 nm moir{\'e} period), consistent with the target angle of 3.6$^{\circ}$. These two insulating states at integer fillings were also observed at lower doping densities in another device D2 with a smaller twist angle of 2.9$^{\circ}$ (Extended Figure 6). The insulating states observed at $\nu=-1$ and $-2$ clearly have a correlated nature, since in the single particle picture, they correspond to a quarter or half-filled honeycomb lattice with the Fermi level lying above or at the Dirac point, respectively\cite{angeli2021gamma,kaushal2022magnetic}. The charge gap opens only when a particular filling is reached. According to our unrestricted Hartree-Fock calculations projected onto the two topmost moir{\'e} bands (Supplementary Information S1), the $\nu=-1$ state is a sublattice-polarized electronic crystal driven by the nearest-neighbor Coulomb repulsion ($U_1$) while the $\nu=-2$ state is a Mott insulator arising from the on-site Coulomb repulsion ($U_0$). We will provide more experimental evidence regarding the nature of these states later.

Next, we plot the oscillator strength $R_{2s}$ as a function of $\nu$ to reveal additional insulating states (Figure 2b). The filling numbers are assigned by assuming a linear dependence between the gate voltage and doping density, based on the equal voltage spacing between the two integer fillings and the charge-neutral condition. In the low-doping density regime, the combined oscillator strength of the IRP and IAP branches is denoted as $R_{2s}$. For $0>\nu>-1$, a series of insulating states appears at commensurate fractional fillings, including $\nu=-\frac{1}{6},-\frac{1}{4}, -\frac{1}{3},-\frac{1}{2},-\frac{2}{3}$, and $-\frac{5}{6}$. These incompressible states are largely insensitive to the external electric field (Figure 2c) because the wavefunction at $\Gamma$-valley VBM is strongly layer hybridized, with an expected coupling strength on the order of hundreds meV. However, some of these states ($\nu\leq-\frac{1}{2}$ and $\nu=-2$) and the phase boundary ($\nu=0$) exhibit dispersion with the electric field, which we attribute to the nonlinear gating effect associated with finite contact and gate resistances. 

At low doping ($\nu\leq-\frac{1}{3}$), the dispersion with a negative slope is known to arise from the high contact resistance between the t-MoSe$_2$ and the electrode\cite{gu2024remote}. The carrier doping density within the sample is controlled by an equivalent gate voltage $V_g$, defined as $V_g=\frac{V_{tg}+1.15V_{bg}} {2}$ (Methods).  As the doping density in the sample increases $-\frac{3}{2}<\nu<-\frac{1}{2}$, the nonlinear gating effect is reduced, and the dispersive feature is eliminated at $\nu=-1$. Although the contact resistance is further reduced as the doping density increases to $ 10^{13}$ cm$^{-2}$, the dispersion reappears near $\nu=-2$, which we attribute to the reduced tunneling resistance of the gate dielectrics when they are operated close to the breakdown voltage. (A circuit model is discussed in Supplementary Information S4) The dispersion becomes much larger if the contact gate voltage is turned off (Figure S4(a)), consistent with the assigned origin of these dispersions. In the meantime, the $\Gamma$ valley character of doped holes is also supported by the minimal electric-field dependence of $R_{2s}$, which varies by less than 10\% for $\nu=-1$ and $-\frac{2}{3}$ at the strongest field limit (Figure 2d). The reduction in $R_{2s}$ at $\nu=-2$ around $E=-50$ mV/nm is due to the doped holes starting to leak into the sensor layer (Figure 2c). Finally, both our Density Functional Theory (DFT) calculations and experimental observation of the indirect exciton photoluminescence (Figure S4(b)) confirm that the VBM in the $\Gamma$ valley is about 200 meV higher than that in the K valley in the extended Brillouin zone of our sample.

\subsection*{Melting of the insulating states}

As the temperature increases, the enhancement of the 2s exciton resonance at both integer and fraction fillings gradually disappears (Figure 3a), indicating the melting of these insulating phases. Since the 2s exciton linewidth also broadens at elevated temperatures, we quantify the melting behavior by plotting the 2s exciton spectral weight contrast, which is the difference in the integrated spectral area around 2s exciton between the insulating and conductive states (Figure S6). The melting temperature, $T_m$, defined as the temperature at which the 2s spectral weight contrast decreases to 10\% of its maximum value\cite{xu2020correlated}, is about 65 K for the $\nu=-2$ state and ranges from 13 to 25 K for the fractional fillings (Figure 3b and Figure 3c). (Additional temperature-dependent data is provided in Extended Figure 3.) Unlike $\nu=-2$, the spectral weight contrast for $\nu=-1$ does not decrease asymptotically with temperature - it first increases upon heating, reaches a maximum at $T^{*}$ ($\sim 15$ K), and then decreases and disappears at about 65 K. This non-monotonic dependence was repeated in another device (D3) with a 4.0$^\circ$ twist angle (Figure S7), resembling the Pomeranchuk effect reported in other TMD moir{\'e} systems\cite{zhang2022tuning}. The $T_m$ at two integer fillings in devices D1 to D3 is summarized in Figure 3d. Interestingly, as the twist angle is reduced from 3.8$^\circ$ to 2.9$^\circ$, the melting temperature of the $\nu=-1$ state shows little variation, while that of the $\nu=-2$ state nearly doubles.     

The thermal melting of the correlated insulating states is known to be governed by two primary mechanisms\cite{erkensten2024stability}. First, when the thermal energy becomes comparable to the size of the charge gap ($\Delta$), the insulating state gradually transitions into a conductive state. Second, phonons can scatter electrons from their equilibrium positions at elevated temperatures. When the deviation exceeds a certain fraction of the interparticle distance, known as the Lindemann criterion, the electronic crystal melts in a way similar to an ionic crystal\cite{goldoni1996stability}. Due to the interplay between these two mechanisms, as well as the potential loss of exciton sensitivities at high temperatures, the $T_m$ measured by 2s exciton sensing provides a lower bound for $\Delta$. The relatively low melting temperature of the insulating states at fractional fillings therefore suggests a smaller $\Delta$ in these states, which is consistent with the long-range nature of the Coulomb repulsion behind the generalized Wigner crystals. On the other hand, the similar $T_m$ values among three devices at $\nu=-1$ suggests that the melting process at this filling is more likely dependent on the local moir{\'e} potential rather than the moir{\'e} period\cite{erkensten2024stability}.

\subsection*{Charge distribution at $\nu=-2$ } 

Unlike the fractional filling and $\nu=-1$ states, the charge distribution in the $\nu=-2$ state is the result of energy competition between the on-site Coulomb repulsion $U_0$ and nearest-neighbor Coulomb repulsion $U_1$\cite{pan2023realizing}. Depending on the ratio between $U_0$ and $U_1$, the $\nu=-2$ state can be either a Mott insulator where each moire site (MX/XM) is filled with one doped hole, or a charge density wave (CDW) state where two holes occupy the same moir{\'e} site, leaving the other site empty (Extended Figure 4a and 4b). Although both scenarios open a charge gap at the Dirac point, the Mott insulator phase breaks the chiral symmetry, while the CDW phase breaks the AB sublattice symmetry. Experimentally, we can distinguish between these two cases by analyzing the doping dependence of moir{\'e} trions in t-MoSe$_2$.

The doping-dependent reflection contrast spectra near the t-MoSe$_2$ 1s exciton ($\sim$ 1.610 eV) in device D1 are presented in Figure 4a. Upon hole doping, the oscillator strength of the charge-neutral exciton gradually decreases, while two new peaks emerge on the low energy side. We label these peaks as $X_1^{+}$ and $X_2^{+}$ since they are similar to the trion in monolayer MoSe$_2$, except that the hole is doped in the $\Gamma$ valley in the twisted bilayer\cite{huang2023quantum}. $X_1^{+}$ has a smaller binding energy ($\sim$17 meV) and reaches the maximum amplitude at $V_{tg}\sim2.67$ V, while $X_2^{+}$ has a larger binding energy ($\sim$30 meV) and maximizes at a higher doping density ($V_{tg}\sim3.50$ V). Interestingly, the doping level $X_1^{+}$ and $X_2^{+}$ reach their maximum is well matched with the conditions $\nu=-1$ and $\nu=-2$, respectively. Although device D2 (having a twist angle $\sim$2.9$^\circ$) has a different set of doping densities for integer fillings, the same correlation is observed (Figure 4b and Extended Figure 7). The $X_2^{+}$ absorption in D2 is slightly enhanced at $\nu=-1$, which we attribute to the reduction in screening - its amplitude is much smaller than that at $\nu=-2$.

We attribute such significant correlations to the spatial character of moir{\'e} trions and the charge distribution in the correlated insulation states. According to our DFT calculations, the excitons and holes in both $X_1^{+}$ and $X_2^{+}$ are located at the XM or MX moir{\'e} site, and the binding energy difference between them likely arises from the distance between exciton and hole (Figure 4c). With this picture, we expect $X_2^{+}$ to be a common tightly bound moir{\'e} trion, where the exciton and hole are located at the same moir{\'e} site, while $X_1^{+}$ is a recently discovered 'charge transfer' trion\cite{liu2024distinct}, where the exciton and hole occupy two different sites, respectively. Since the exciton-hole distance is larger in $X_2^{+}$, the Coulomb attraction and binding energy become smaller. 

Following these assignments, we can explain the doping dependence of the two moir{\'e} trions. At $\nu=-1$, the oscillator strength of $X_1^{+}$ peaks because half of the MX/XM sites are filled, maximizing the chance for an exciton in an unfilled site to bind with a doped hole in the neighboring site\cite{ciorciaro2023kinetic}. When the filling increases to $\nu=-2$, the fact that $X_2^{+}$ reaches a maximum means the doped holes occupy every MX and XM site so that each exciton can bind with a doped hole in the same moir{\'e} potential well, forming a moir{\'e} trion. As the doping level increases further, the holes at a doubly occupied site can no longer couple to any exciton, thus reducing the amplitude of $X_2^+$ trion\cite{ciorciaro2023kinetic}. Therefore, these observations strongly suggest that the $\nu=-2$ state is a Mott insulator where each sublattice is filled with one hole.

Given the Mott nature of the insulating state at $\nu=-2$, we can estimate the lower bound of the on-site Coulomb repulsion energy $U_0$ based on its melting temperature $T_m$. According to numerical results in the literature\cite{tang2013finite}, $U_{0}\approx4.75k_{B}T_{m}$, which corresponds to about 25 meV in device D1. On the other hand, the nearest-neighbor hopping integral $t$ is estimated to be about 5 meV in D1, following the MacDonald-Bistritzer model\cite{angeli2021gamma}. The resulting $U_0/t$ ratio is therefore 5, larger than the critical value of 4 for the semimetal-insulator transition in a honeycomb lattice\cite{ostmeyer2020semimetal}. A similar calculation shows that device D2 has a lower bound for $U_0/t$ of 17, which is also consistent with the correlated insulating phase we observed. In addition, comparing the retrieved $U_{0}$ with our theoretical model, we obtain an effective dielectric constant $\epsilon\approx 27-30$. The Hartree-Fock calculations under such a condition agree with the observation that $T_m$ of $\nu=-2$ decreases with increasing twist angle, indicating that the semimetal-insulator phase transition is expected to occur at a larger twist angle or under stronger screening conditions (Extended Figure 4).

\subsection*{Spin-spin coupling at integer fillings}

Finally, we use the moir{\'e} trions in t-MoSe$_2$ to explore the magnetic properties of correlated insulating states at integer fillings. As shown in Figure 5a, we first measure the magnetic circular dichroism (MCD) of D1 at different photon energies under an out-of-plane magnetic field at both $\nu=-1$ and $-2$. At $\nu=-1$, $X_2^{+}$ exhibits an MCD signal that is proportional to the magnetic field. While $X_1^{+}$ shows a negligible MCD signal, a peak near 1.620 eV shows a similar MCD signal with an opposite sign. In contrast, we do not observe any significant MCD signal at $\nu=-2$ even under the largest magnetic field applied (8 T). 

We attribute the MCD signal in $X_2^{+}$ to the intervalley exchange interaction between the doped holes in the $\Gamma$ valley and excitons in the $K$-valley, which lifts the energy degeneracy between the spin-singlet and spin-triplet trions (Figure 5b). Due to the large spin-orbital coupling at the Brillouin zone corners, an optically excited bright exciton in the $\pm K$ valley is composed of a hole of either up- or downward spin, which preferentially binds with an additional hole of opposite spin in the $\Gamma$ valley. This intervalley spin-triplet trion preserves the optical selection rule in TMD monolayers, meaning that a finite MCD signal corresponds to a spin polarization among the doped holes. Although the exchange interaction in $X_2^{+}$ can be on the meV scale, we expect it to be much weaker in $X_1^{+}$, where the exciton is spatially separated from the hole, resulting in a negligible MCD signal\cite{naik2022intralayer}. The high-energy peak is attributed to the repulsive Fermi polaron, which has been reported to exhibit the opposite optical selection rule compared to the attractive branch (trion)\cite{tao2024valley}.

If we extend this picture to the $\nu=-2$ state, the absence of an MCD signal suggests a zero magnetic susceptibility, indicating that the insulating state may be an antiferromagnetic correlated state with a characteristic temperature well above the measurement temperature. Alternatively, theoretical investigations have proposed a gapped spin liquid (a short-range resonating valence bond state) in a half-filled honeycomb Hubbard model near the $U_0/t$ condition for device D1\cite{meng2010quantum}. Within this framework, our negligible MCD signal indicates the spin gap is larger than the Zeeman or thermal energy. Experimentally, we cannot distinguish between these scenarios or other possibilities within the limited temperature range in our experiments. Because of the small exchange energy in moir{\'e} trions, the MCD signal no longer reflects the spin polarization of doped carriers at temperatures higher than 7 K, as indicated by the deviation from the Curie-Weiss (CW) law in Figure 5d.

Finally, we further examine the spin-spin coupling among doped holes at $\nu=-1$ by measuring the magnetic field dependence of the MCD signal at various temperatures (Figure 5c). At base temperature ($T$ = 1.6 K), MCD increases linearly with the magnetic field until it becomes saturated at $|B|=3.8 \pm 0.2$ T (the MCD signal is integrated within a 5-meV window centered around 1.620 eV). At higher temperatures, the linear slope decreases, and the saturation field becomes larger, a trend previously observed in TMD heterobilayers \cite{tang2020simulation,xu2020correlated}. The magnetic susceptibility is extracted from the slope near the zero field ($\chi_{zz}=\lim_{B\to0}\frac{\partial M}{\partial B}$), which follows a linear CW law with a small positive CW temperature $\Theta_{CW}=+0.20 (\pm 0.20)$ K (Figure 5d). The small yet finite CW temperature is confirmed by measurements at another point (Extended Figure 5, $\Theta_{CW}=+1.10 (\pm 0.40)$). The MCD's magnetic field dependence becomes indistinguishable above 7 K, likely because of the limited exchange effect and spin sensitivity of moir{\'e} trions.

The small positive $\Theta_{CW}$ in our fitting suggests a weak ferromagnetic interaction between the local moments at $\nu=-1$, indicating a direct-exchange interaction larger than the antiferromagnetic superexchange interaction, which is consistent with our theoretical calculations (details can be found in the Supplementary Information S3). By including perturbations up to the fourth order in a Heisenberg model (Extended Figure 5), we theoretically find that the ferromagnetic spin-spin coupling $J$ varies with the screening strength but is always negative in our interest range (see the inset of Figure 5d). These results highlight the importance of nonlocal Coulomb interactions in determining the magnetic ground state of correlated insulating states in a honeycomb lattice\cite{morales2022nonlocal}.

\subsection*{Conclusion}

In conclusion, we have observed a range of correlated insulating states at integer and fractional fillings in twisted MoSe$_2$ homobilayers arising from the $\Gamma$-valley moir{\'e} flat bands. Rather than forming a semimetal, the half-filled honeycomb lattice in our system is a Mott insulator with antiferromagnetic coupling between carriers. The quarter-filled state is a generalized Wigner crystal with weak ferromagnetic coupling and a melting mechanism distinct from that of the Mott state. The strong correlation effect persists up to 110 K and is enabled by the substantially reduced Fermi velocity in our system. The moir{\'e} trions with different spatial characters and 2s exciton sensing techniques establish a powerful optical approach to probe the electronic compressibility, charge distribution, and spin-spin coupling among carriers doped in the $\Gamma$ valley. The absence of a magnetic response at half filling points to an unconventional magnetic ground state, making its definitive identification a crucial step in understanding strongly correlated fermions on a honeycomb moir{\'e} superlattice.

\textbf{Note:} While preparing this manuscript, we notice that a work on the same topic appears on Arxiv (\textbf{arXiv: 2412.07150}) from the Cornell group.

\section*{Methods}
\subsection*{Sample Fabrication}

The twisted MoSe$_2$ moir{\'e} device is fabricated using the standard dry transfer method under ambient conditions\cite{wang2013one}. Thin-layer hBN flakes, few-layer graphene, monolayer MoSe$_2$, and monolayer WS$_2$ are exfoliated on 285 nm SiO$_2$/Si substrate. The flake thicknesses are confirmed by atomic force microscopy (AFM) and optical contrast. A 1030 nm ultrafast laser is used to cut the few-layer graphite into specific shapes for gate electrodes. A thin film of polycarbonate, placed on polydimethylsiloxane (PDMS), is used as a handle to pick up layers in the following sequence: top hBN, top graphite, top gate hBN, graphene contact, monolayer WS$_2$ sensor, $\sim$1 nm hBN spacer, twisted MoSe$_2$ bilayer, bottom gate hBN, bottom graphite, and contact gate graphite. The monolayer MoSe$_2$ is cut into two pieces by an ultrafast laser at 750 nm. Half of the MoSe$_2$ monolayer is picked up first, and the other half is rotated by a target angle and then picked up at 60-100 \textcelsius. The pick-up temperature for other flakes is 100 \textcelsius. Finally, the device is released on a Si/SiO$_2$ substrate with pre-patterned gold electrodes to form contact at 180 \textcelsius. The polycarbonate residue is washed off in chloroform for 30 minutes and cleaned up in IPA for 5 minutes. 

\subsection*{Optical measurements}

All optical measurements, including reflection contrast spectroscopy and magnetic circular dichroism (MCD), are performed in a close-cycle cryostat (Attodry-2100, base temperature 1.6 K) equipped with a superconducting magnet. A spatially filtered broadband tungsten halogen lamp is focused onto the device with a spot size of a few micrometers via a low-temperature objective lens (NA = 0.65). The intensity was kept below 10 nW/$\mu$m$^2$ to prevent sample heating. The reflected light is spectrally resolved using a spectrometer (Princeton Instruments) with a thermal-electric-cooled CCD camera. To obtain the reflection contrast ($\Delta R/R$) spectrum, we take a reference spectrum at the position outside the sample area. To extract the oscillator strength of the 2s exciton, we subtract a slowly varying background due to multilayer reflections, using third-order polynomial fitting in the photon energy range displayed in Extended Figure 2. Then, we use the peak-to-peak amplitude to represent the oscillator strength.  

For MCD measurements, we used a combination of a linear polarizer and an achromatic quarter-wave plate to generate left-handed ($\sigma^+$) and right-handed ($\sigma^-$) circularly polarized light. The MCD spectrum is defined as the difference between the left- and right-handed reflected light intensity: $MCD=\frac{R_{\sigma^+}-R_{\sigma^-}}{R_{\sigma^+}+R_{\sigma^-}}$. The reflectance spectrum for $\sigma^+$ and $\sigma^-$ light is acquired by using the reflected light spectrum of the sample with the heavily hole-doped condition as a reference. We integrate the MCD signal over a spectral window of 5 meV near the Fermi-polaron ($RP$) resonance. Magnetic susceptibility is extracted by linear fitting of the MCD response at low magnetic fields ($|B|\leq 1.2$ T).        

\subsection*{Twist angle calibration and electrostatic model}
We use the gate voltage of two integer fillings ($\nu=-1$ and $\nu=-2$) as benchmarks to calibrate the twist angle $\theta$. The moir{\'e} density $n_M$ is determined by the voltage differences between the top and bottom gates ($\Delta V_{tg}$ and $\Delta V_{bg}$) between two integer fillings: 
\begin{align}
    n_M=\frac{1}{e}(C_{tg} \Delta V_{tg} +C_{bg} \Delta V_{bg}) \notag
\end{align}
$C_{tg}$ and $C_{bg}$ are the geometrical capacitance of top and bottom gate.  
\begin{align}
    C_{tg}&=\frac{\epsilon_0}{d_t/\epsilon_{hBN}+d_0/\epsilon_{WS_2}+d_s/\epsilon_{hBN}} \notag \\
    C_{bg}&=\frac{\epsilon_{hBN}\epsilon_0}{d_b} \notag
\end{align}
The relative permittivities are $\epsilon_{hBN}\approx 3$ and $\epsilon_{WS_2} \approx 7.5$ for hBN and WS$_2$ sensor layer. The thicknesses of the top and bottom hBN layers are $d_t \approx 6.6$ nm and $d_b \approx 9.5$ nm, respectively. (confirmed by AFM in contact mode). The spacer hBN thickness $d_s$ is 1 nm and the sensor thickness $d_0 \approx 0.65$ nm. For the position P1 examined in the main text, the moir{\'e} density of $n_M$ is $4.6\times 10^{12}$ cm$^{-2}$, corresponding to a twist angle of $\theta=\frac{180}{\pi}a\sqrt{\frac{\sqrt{3}}{2} n_M} \approx 3.8^\circ$, where $a\approx 0.33$ nm is the lattice constant of the MoSe$_2$ monolayer. The electric field ($E$) within the MoSe$_2$ bilayer is $E=\frac{1}{2}(\frac{V_{tg}}{d_t+d_0+d_s}-\frac{V_{bg}}{d_b})$. To maintain a zero electric field condition, we set the ratio between $V_{bg}$ and $V_{tg}$ to be 1.15, as shown in Figure 2.

\section*{Data availability} 
All data needed to evaluate the conclusions in the paper are present in the paper. Additional data are available from the corresponding authors upon reasonable request.

\section*{Acknowledgement} 
Z.Y., D.Y., J.L. H.H., J.D., M.Z., N.K. acknowledge support from the Natural Sciences and Engineering Research Council of Canada, Canada Foundation for Innovation, New Frontiers in Research Fund, Canada First Research Excellence Fund, Max Planck-UBC-UTokyo Center for Quantum Materials, and Gordon and Betty Moore Foundation's EPiQS Initiative (Grant GBMF11071). Z.Y. is also supported by the Canada Research Chairs Program. K.W. and T.T. acknowledge support from JSPS KAKENHI (Grant Numbers 19H05790, 20H00354, and 21H05233). C. E. H. acknowledges the funding support of the National Science and Technology Council, Taiwan, under the Grant No. 113-2917-I-032001. Z. L. acknowledges the support from the U.S. National Science Foundation under Grant No. DMR- 2440763.

\section*{Author Contributions} 
Z.Y. conceived the project. D.Y., J.L., and H.H. fabricated the devices and performed all the optical measurements with the assistance of J.D.. D.Y. and J.L. built the experimental setup. K.W. and T.T. provided the hBN crystal. N.K. performed the theoretical model calculation under the supervision of M.F.. C.H. performed the DFT calculation under the supervision of Z.L. Z.Y. and D.Y. analyzed the data. Z.Y. and D.Y. wrote the manuscript based on the input from all other authors. D.Y., J.L., and H.H. contributed equally to this work. 

\section*{Competing interests}
The authors declare no competing interests.

\section*{Figures}

\begin{figure}[ht]
    \includegraphics[width=1\textwidth]{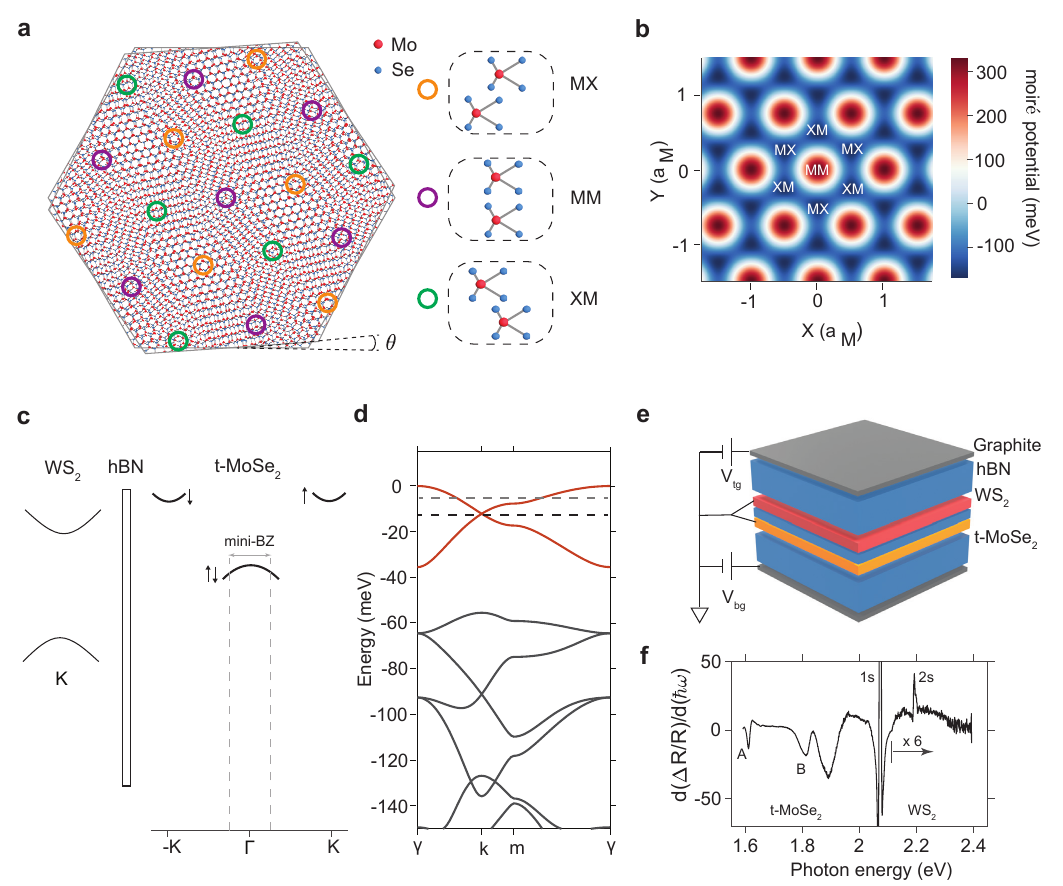}
    \textbf{Figure 1 $|$ Honeycomb moir{\'e} superlattice arising from a folded valence band maximum in the $\Gamma$ valley.} \textbf{a.} Schematics of a small-angle twisted MoSe$_2$. High symmetry points are highlighted by colored circles. The yellow circle represents the MX sites where the Mo atom is on top of the Se atom, while the green circle and purple circle correspond to XM and MM sites, respectively. \textbf{b.} Calculated moir{\'e} potential for doped holes. \textbf{c.} Type-II band alignment between t-MoSe$_2$ and WS$_2$ as well as the moir{\'e} band folding in the $\Gamma$ valley.  The two dashed lines denote the moir{\'e} Brillouin zone (moir{\'e} BZ) boundary. The $\Gamma$-valley band is spin-degenerate, while the bands in $K$($-K$) valley are spin-polarized due to large spin-orbital coupling. \textbf{d.} Calculated band structure of t-MoSe$_2$ bilayer along the direction of $\gamma$-k-m-$\gamma$ in the moire BZ. Grey and black dashed lines denote the Fermi level when the filling is at $\nu=-1$ and $\nu=-2$. \textbf{e.} Schematic of a dual-gated device with a WS$_2$ sensing layer. \textbf{f.} The first energy derivative of the reflection contrast spectrum of the sample at the intrinsic condition.
    \label{fig: Fig.1}
\end{figure}

\begin{figure}
    \includegraphics[width=1\textwidth]{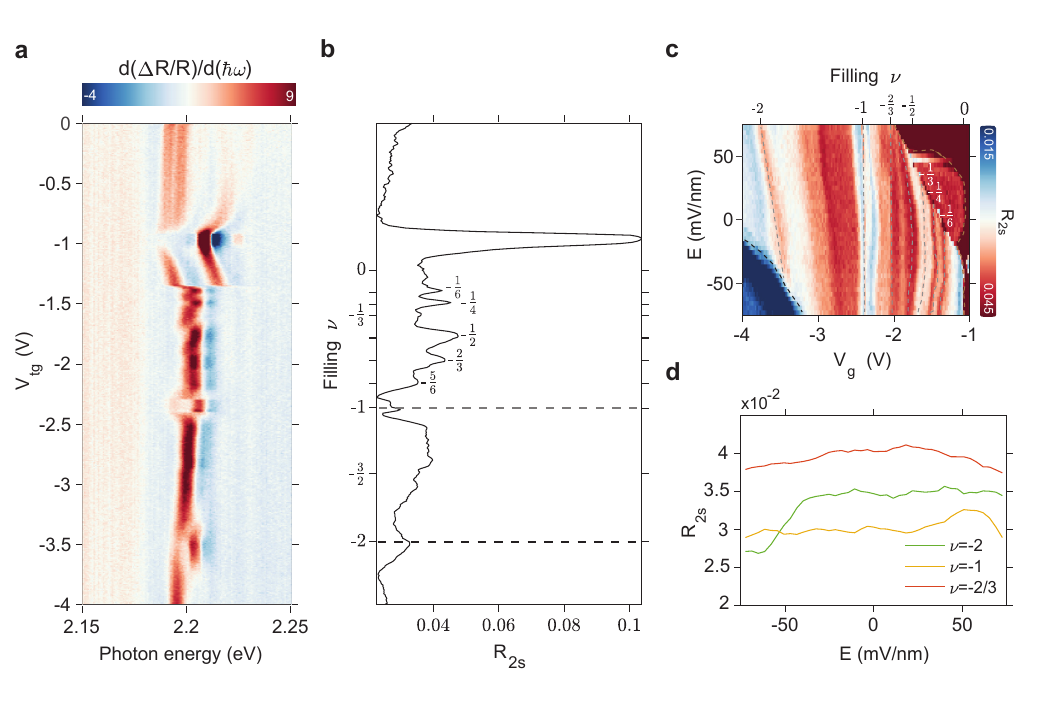}
    \textbf{Figure 2 $|$ Correlated insulating states at integer and fractional fillings.} \textbf{a.} The first energy derivative of reflection contrast of the device near the 2s exciton resonance versus top gate voltage. The backgate voltage is applied with a ratio of $V_{bg}=1.15V_{tg}$. The experimental temperature is 7 K. \textbf{b.} Filling number ($\nu$) versus oscillator strength of the 2s exciton ($R_{2s}$). Grey and black dashed lines label the two integer fillings at $\nu=-1$ and $\nu=-2$, respectively. \textbf{c.} The phase diagram of $R_{2s}$ as a function of doping controlled by the effective gate voltage $V_g=\frac{V_{tg}+1.15V_{bg}} {2}$ and electric field ($E$). The phase boundary is labeled by a brown dashed line.  The insulating states are traced by grey dashed lines with the corresponding filling numbers labeled. The black dashed line indicates the condition where the doped carriers start to leak into the sensor layer. \textbf{d.} $R_{2s}$ as a function of the electric field at three representative fillings.
    \label{fig: Fig.2}
\end{figure}

\begin{figure}
    \includegraphics[width=1\textwidth]{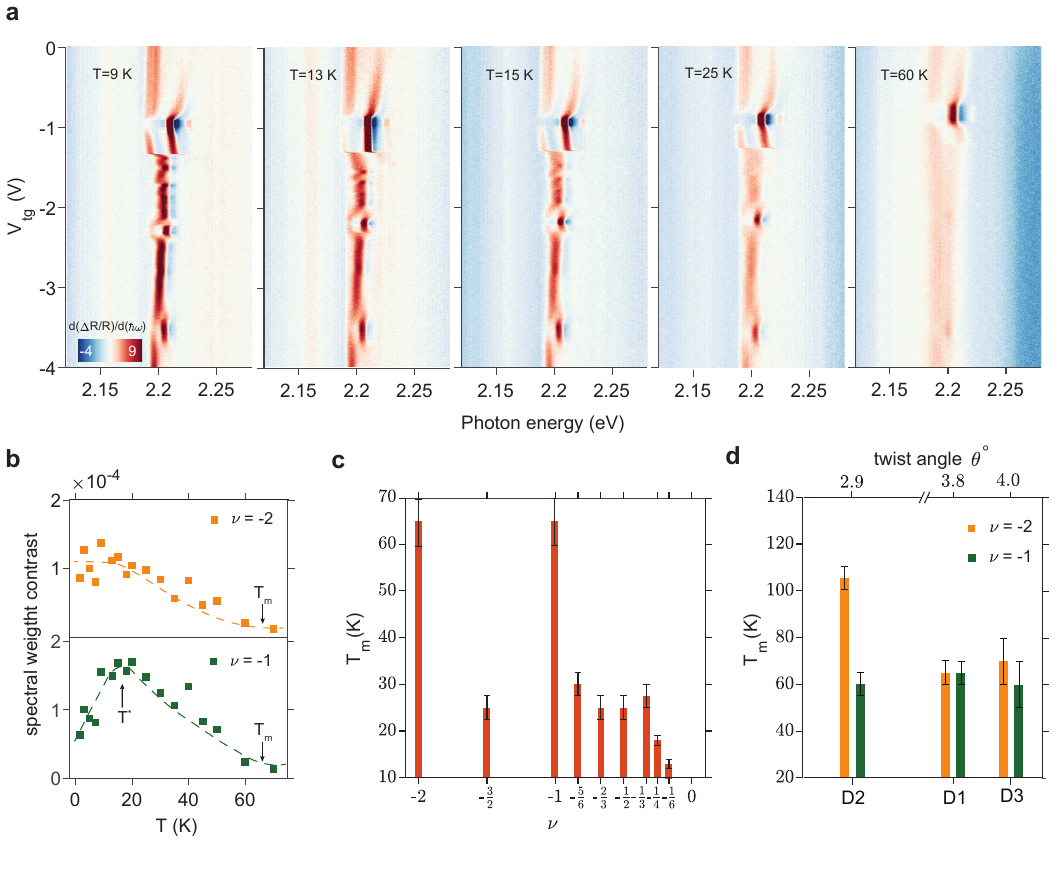}
    \textbf{Figure 3 $|$ Temperature dependence of the correlated insulating states.} \textbf{a.} The first energy derivative of reflection contrast as a function of $V_{tg}$ measured at different temperatures. \textbf{b.} Temperature dependence of the 2s exciton spectral weight contrast for the insulating states at $\nu=-1$ (green) and $\nu=-2$ (yellow). The dashed lines are guides to the eye. $T_m$ denotes the melting temperature. $T^* \approx 15$ K denotes the temperature where the 2s spectral weight contrast exhibits a peak. \textbf{c.} The melting temperature of all observed insulating states at both integer and fractional fillings. \textbf{d.} Combined plot of  $T_m$ for $\nu=-1$ (green) and $\nu=-2$ (yellow) states at three different twist angles.
    \label{fig: Fig.3}
\end{figure}

\begin{figure}
    \includegraphics[width=1\textwidth]{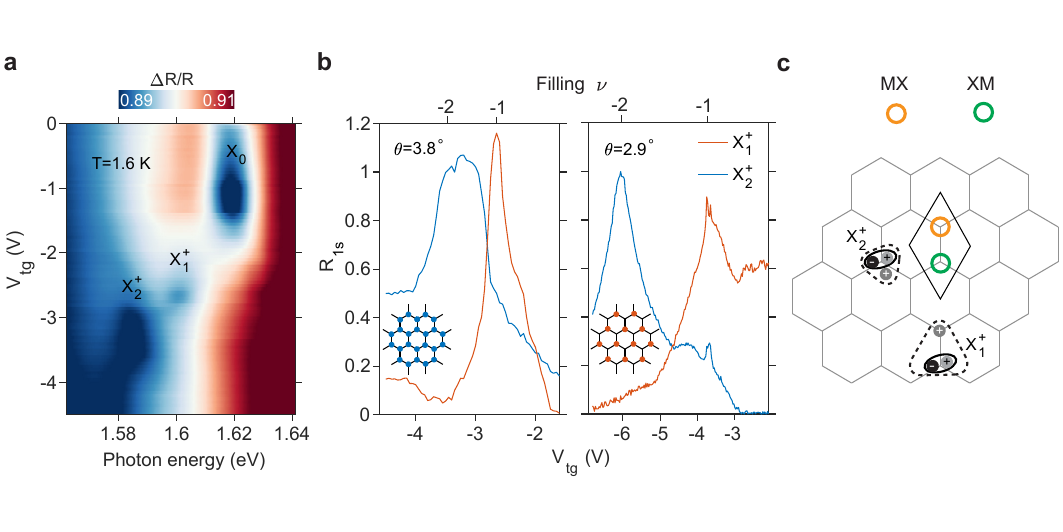}
    \textbf{Figure 4 $|$ The charge distribution of $\nu=-2$ state.} \textbf{a.} The reflection contrast spectrum as a function of gate voltage near the A exciton resonance of t-MoSe$_2$ in device D1. $X_2^{+}$ and $X_1^{+}$ denote the tightly bound and charge transfer moir{\'e} trions, respectively. $X_0$ is the charge-neutral exciton. \textbf{b.} Oscillator strength of $X_1^{+}$ (red) and $X_2^{+}$ (blue) as a function of doping for devices D1 (left) and D2 (right). The two insets plot the charge distribution of the insulating states at $\nu=-2$ (left) and $\nu=-1$ (right). \textbf{c.} Schematics for tightly bound ($X_2^{+}$) and charge transfer ($X_1^{+}$) moir{\'e} trions. 
    \label{fig: Fig.4}
\end{figure}

\begin{figure}
    \includegraphics[width=1\textwidth]{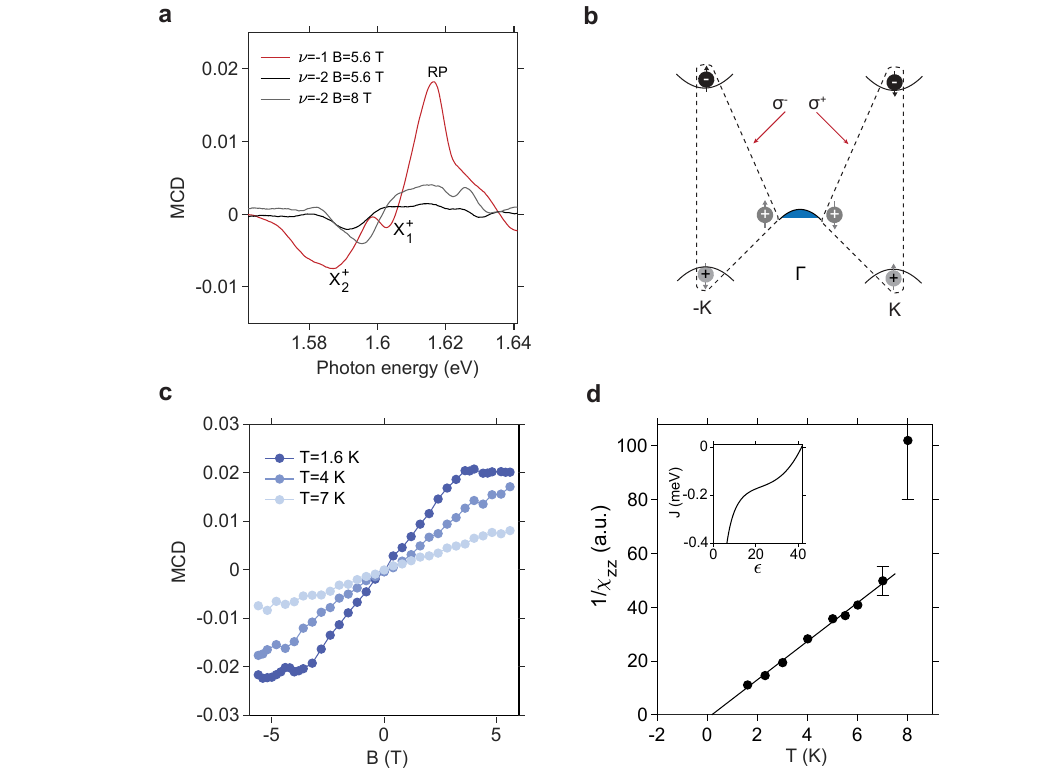}
    \textbf{Figure 5 $|$ Magnetic properties of the insulating states at integer fillings.} \textbf{a}. Magnetic circular dichroism (MCD) versus photon energy for $\nu=-1$ at B=5.6 T and T=1.6 K (red), $\nu=-2$ at B=5.6 T and T=1.6 K (black) and at B=8 T and T=1.6 K (grey). $X_1^{+}$, $X_2^{+}$, and $RP$ denote the two moir{\'e} trions and one repulsive Fermi polaron, respectively. \textbf{b.} A schematic illustration of the MCD mechanism. The $\pm K$-valley excitons couple with the holes doped in the $\Gamma$ valley, forming $\Gamma$-K trions (Fermi polarons) with opposite optical selection rules. \textbf{c.} Magnetic field dependence of MCD at different temperatures. The doping density is set to $\nu=-1$. The MCD is integrated near the resonance of $X_2^{+}$. \textbf{d}. Curie-Weiss fit for the magnetic susceptibility ($\chi_{zz}$) near zero magnetic field as a function of temperature at $\nu=-1$. The fitted Curie-Weiss temperature is $\Theta_{CW}=+0.20(\pm 0.20)$ K. The uncertainty arises from the linear fitting. \textbf{Inset:} The spin-spin coupling $J$ of $\nu=-1$ state is calculated by a perturbation theory based on the Heisenberg model.
    \label{fig:Fig.5}
\end{figure}

\begin{figure}
    \includegraphics[width=1\textwidth]{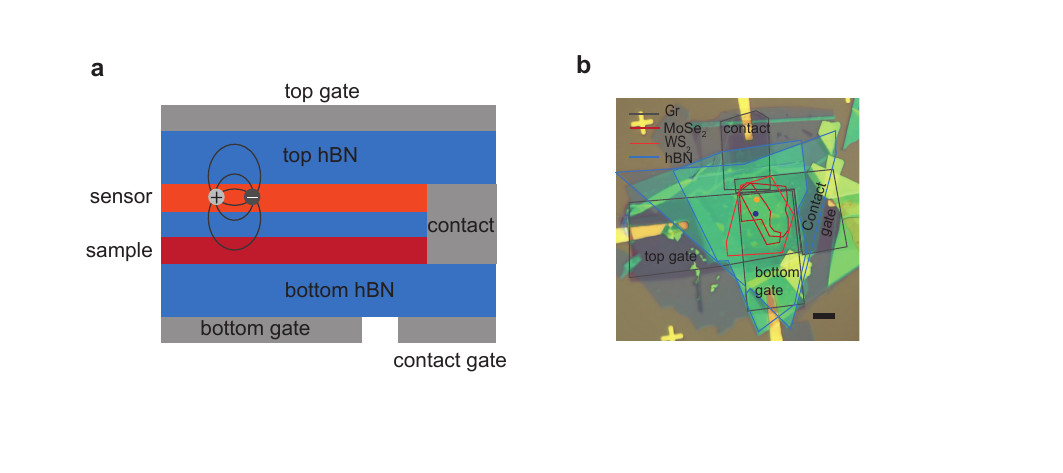}
    \textbf{Extended Figure 1 $|$ Schematic and optical image of device D1.} \textbf{a.} A schematic side view of Device D1. A contact gate covers the TMD-graphene contact region to minimize the nonlinear gating effect. \textbf{b.} The optical image of device D1. The twisted MoSe$_2$ homobilayer is outlined by the solid red curve. The grey curves outline the region of graphite gates. The blue and yellow dots denote the two measured positions P1 and P2 in D1. Scale bar: 5 $\mu$m. 
    \label{fig:EXFig.1}
\end{figure}

\begin{figure}
    \includegraphics[width=1\textwidth]{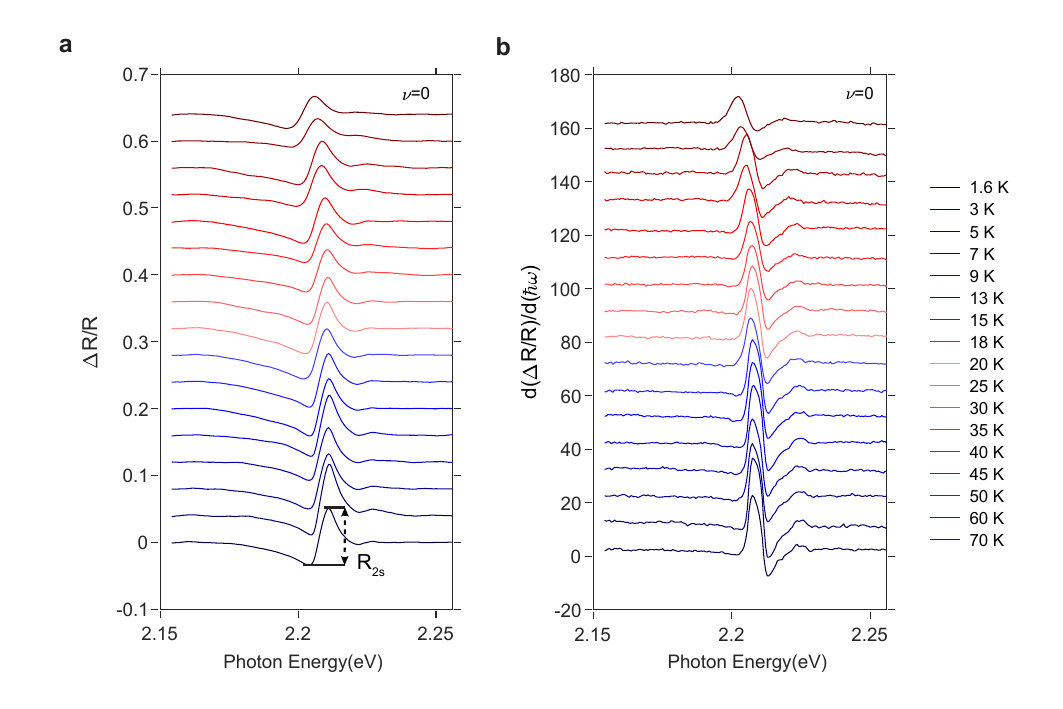}
    \textbf{Extended Figure 2 $|$ Reflection contrast and its first energy derivative spectra of the 2s exciton.} \textbf{a.} Reflection contrast spectra of the 2s exciton at different temperatures. The doping is fixed at the charge-neutral point of $\nu=0$. We use the peak-to-peak amplitude at the 2s resonance to represent the oscillator strength, denoted as $R_{2s}$.  \textbf{b.} The first derivative of the spectra in \textbf{(a)} with respect to photon energy.
    \label{fig:EXFig.2}
\end{figure}

\begin{figure}
    \includegraphics[width=1\textwidth]{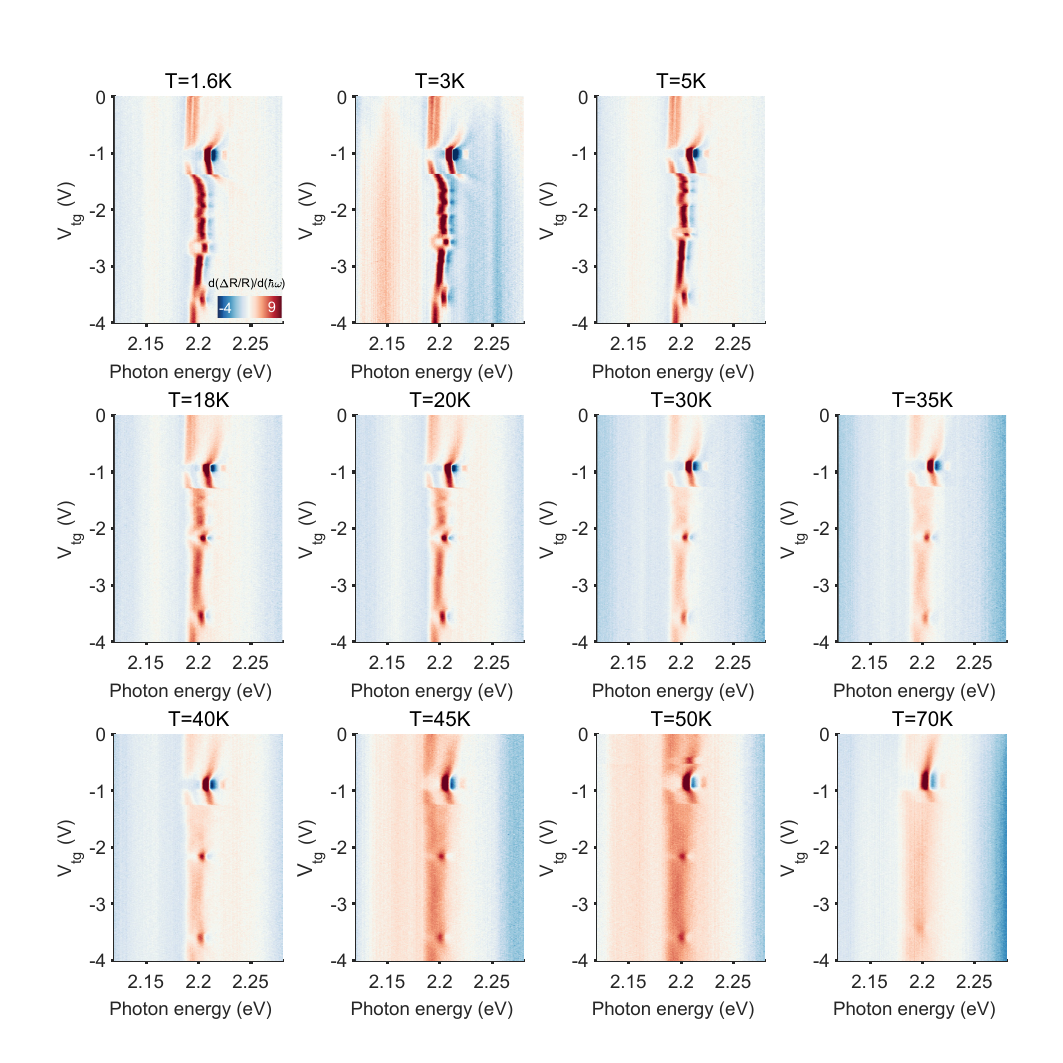}
    \textbf{Extended Figure 3 $|$ Additional temperature dependence data for device D1.} Contour plot of gate dependence data at T=1.6, 3, 5, 18, 20, 30, 35, 40, 45, 50, and 70 K.
    \label{fig:EXFig.3}
\end{figure}

\begin{figure}
    \includegraphics[width=1\textwidth]{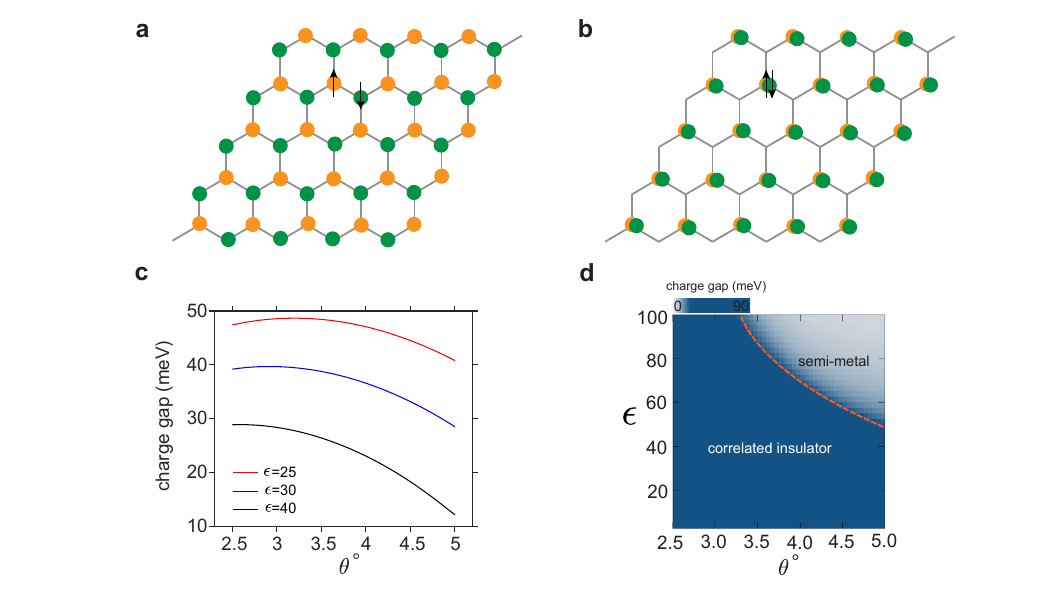}
    \textbf{Extended Figure 4 $|$ Hartree-Fock calculations of the charge gap at $\nu=-2$}  \textbf{a,b.} Schematics of two possible charge configurations for $\nu=-2$ state. When the ratio of $U_0/t$ is large enough, the ground state can be a spin-staggered antiferromagnetic insulator (\textbf{a}). Due to competition between U$_0$ and U$_1$, another possible charge distribution of the ground state is a sublattice-polarized charge density wave state (\textbf{b}). \textbf{c.}  Calculated gap size of $\nu=-2$ state as a function of twist angle for three different effective dielectric permittivity.  \textbf{d.} Charge gap of $\nu=-2$ state as a function of the effective dielectric permittivity ($\epsilon$) and twist angle ($\theta$), calculated using unrestricted Hartree-Fock method.  The dashed orange denotes the boundary between the correlated insulator and semimetal phases.
    \label{fig:EXFig.4}
\end{figure}

\begin{figure}
    \includegraphics[width=1\textwidth]{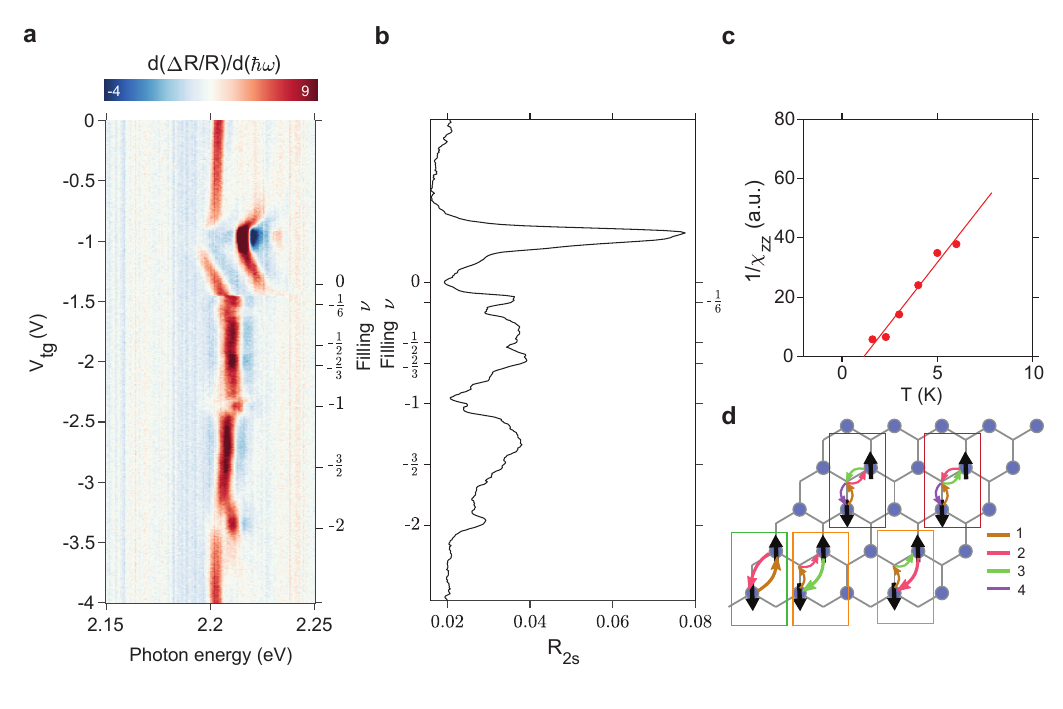}
    \textbf{Extended Figure 5 $|$ Additional data for another position P2 of device D1.} \textbf{a.} First energy derivative of the reflection contrast spectra as a function of top gate voltage. The experimental temperature is 7 K. \textbf{b.} Filling number versus oscillator strength of the 2s exciton at position P2. \textbf{c.} Curie-Weiss fit for the magnetic susceptibility near zero magnetic field as a function of temperature at filling $\nu=-1$ and position P2. \textbf{d.} A schematic of the hopping processes contributing to the leading terms in the spin exchange. Black arrows represent the local moments. Green, yellow, and red boxes outline the second-, third-, and fourth-order processes, respectively. The color of the arrows corresponds to the step number in the hopping process.
    \label{fig: EXFig.5}
\end{figure}

\begin{figure}
    \includegraphics[width=1\textwidth]{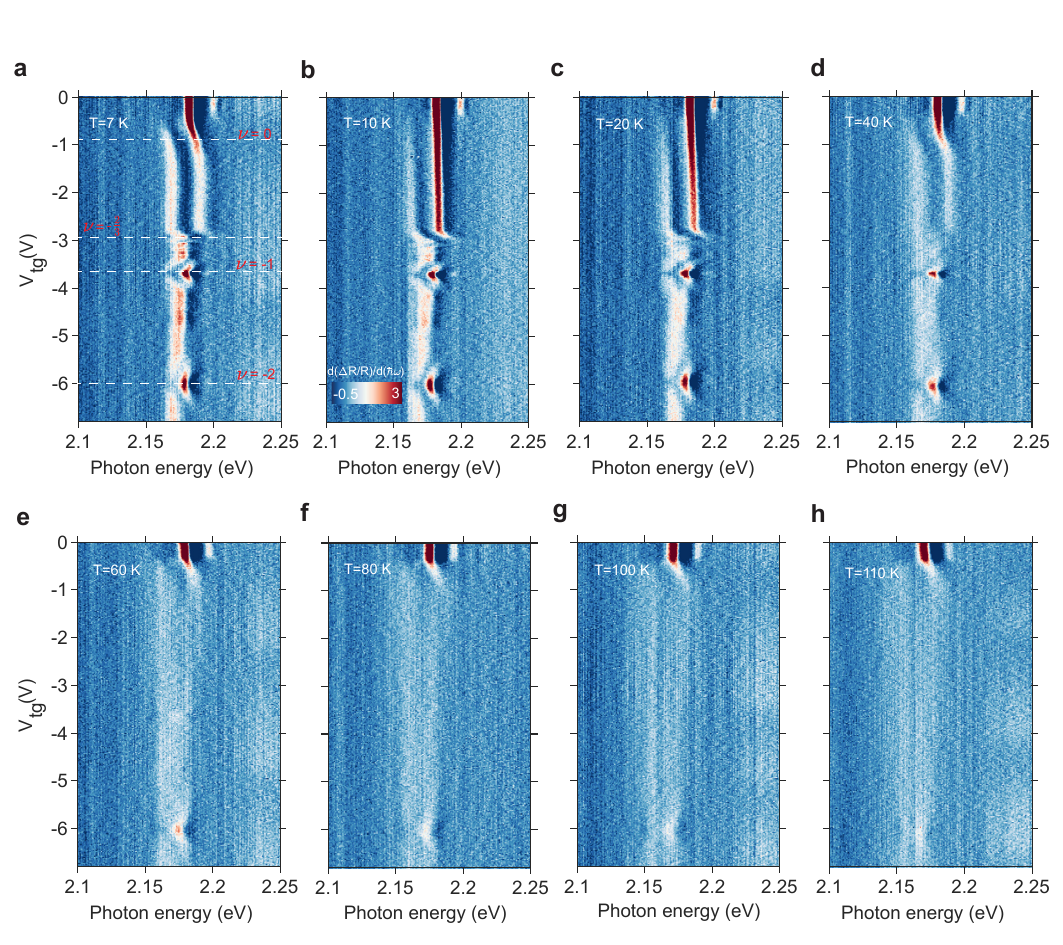}
    \textbf{Extended Figure 6 $|$ Temperature dependence of the insulating states in device D2.} \textbf{a-h} Contour plot of gate-dependent reflection contrast spectra at T=7, 10, 20, 40, 60, 80, 100, and 110 K. The white dashed lines in \textbf{(a)} denote the filling of $\nu=0$,$-\frac{2}{3}$, $-1$, and $-2$ states. The calibrated twist angle for D2 is 2.9$^\circ$.
    \label{fig: EXFig.6}
\end{figure}

\begin{figure}
    \includegraphics[width=1\textwidth]{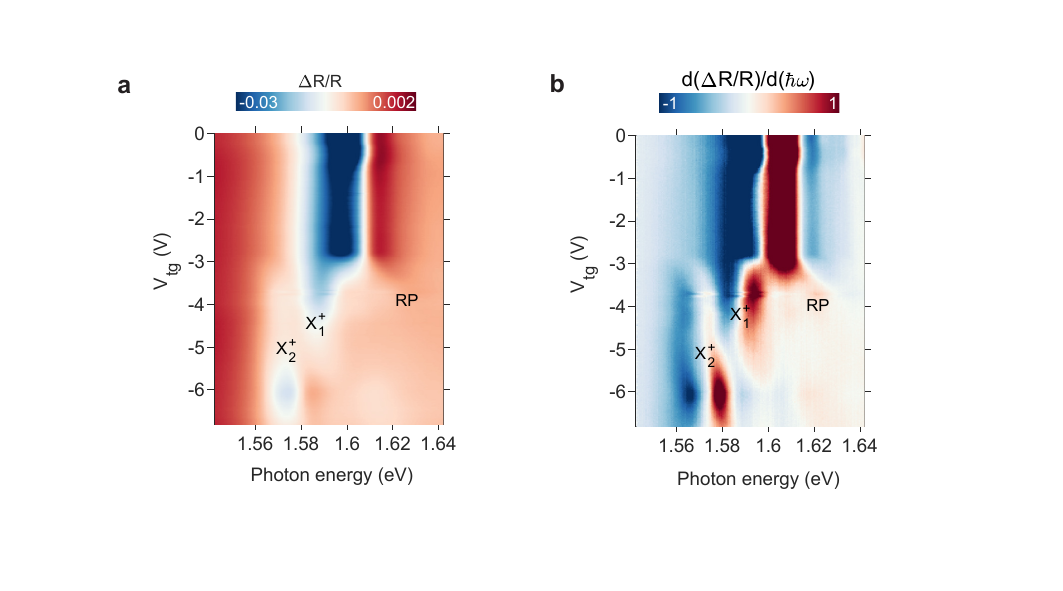}
    \textbf{Extended Figure 7 $|$ The 1s spectrum of t-MoSe$_2$ in Device D2.} \textbf{a.} The reflection contrast spectrum as a function of gate voltage ($V_{tg}$) near the A exciton resonance of t-MoSe$_2$ in device D2. $X_1^{+}$ and $X_2^{+}$ denote the charge transfer and tightly bound trions (attractive polarons), respectively. RP denotes the repulsive branch. \textbf{b.} The first energy derivative of the reflection contrast spectra in \textbf{(a)}.   
    \label{fig: EXFig.7}
\end{figure}

\end{document}